\documentclass[twocolumn,prb,showpacs,aps]{revtex4}

\usepackage{graphicx}
\usepackage{amssymb,amsmath}
\usepackage{times}
\newcommand{\be}{\begin{equation}}
\newcommand{\ee}{  \end{equation}}
\newcommand{\ba}{\begin{eqnarray}}
\newcommand{\ea}{  \end{eqnarray}}
\newcommand{\bi}{\begin{itemize}}
\newcommand{\ei}{  \end{itemize}}
\newcommand{\ve}{\varepsilon}

\begin{document}

\title{Adiabatic Charge Pumping through Quantum Dots in the Coulomb Blockade Regime}

\author{A. R. Hern\'andez}
\affiliation{Laborat\'orio Nacional de Luz S\'{\i}ncrotron,
             Caixa Postal 6192, 13083-970 Campinas, Brazil}
\affiliation{Centro Brasileiro de Pesquisas F\'{\i}sicas,
             Rua Dr.~Xavier Sigaud 150, 22290-180 Rio de Janeiro, Brazil}
\author{F. A. Pinheiro}
\affiliation{Instituto de F\'{\i}sica, Universidade Federal do Rio de Janeiro,
             Caixa Postal 68528, 21945-972 Rio de Janeiro, Brazil}
\author{C. H. Lewenkopf}
\affiliation{Instituto de F\'{\i}sica,
             Universidade do Estado do Rio de Janeiro,
             Rua S\~ao Francisco Xavier 524, 20550-900 Rio de Janeiro, Brazil}

\author{E. R. Mucciolo} \affiliation{Department of Physics, University
  of Central Florida, Orlando, Florida, 32816-2385, USA }

\begin{abstract}

We investigate the influence of the Coulomb interaction on the
adiabatic pumping current through a quantum dot. Using nonequilibrium
Green's functions techniques, we derive a general expression for the
current based on the instantaneous Green's function of the dot. We
apply this formula to study the dependence of the charge pumped per
cycle on the time-dependent pumping potentials. Motivated by recent
experiments, the possibility of charge quantization in the presence of
a finite Coulomb repulsion energy is investigated.

\end{abstract}

\date{\today}


\maketitle


\section{Introduction}
\label{intro}

\noindent

The basic idea of electron pumping, put forward in the pioneer work of
Thouless,~\cite{thouless} is to generate a DC current through a
conductor in the absence of an applied bias voltage. This may be
accomplished by applying time-dependent perturbations to the
conductor. In electronic transport through mesoscopic conductors, the
typical experimental time scale over which these external
perturbations vary is large compared to the lifetime of the electron
inside the conductor (dwell time). In that case, the pumping mechanism
is called adiabatic.

Adiabatic quantum pumping in mesoscopic noninteracting open quantum
dots was investigated theoretically by Brouwer \cite{brouwer98} by
means of a scattering approach. Applying the emissivity theory
introduced by B\"uttiker and co-workers,\cite{buttiker94} he
demonstrated that the pumping current is proportional to the driving
frequency and shows large mesoscopic fluctuations accounted by Random
Matrix Theory. This scattering approach has been employed to
investigate several aspects of adiabatic quantum pumping in
noninteracting systems, such as the role of discrete symmetries on the
pumped charge,\cite{aleiner00} the effects of inelastic scattering and
decoherence,\cite{MoskButt01,cremers02} the role of noise and
dissipation,\cite{MoskButt02} Andreev interference effects in the
presence of superconducting leads,\cite{tadei04,blaauboer02} as well
as spin pumping.\cite{mucciolo02,governale03,moises04,Mucciolo07}
Pumping phenomena in noninteracting systems have also been
investigated using alternative theoretical approaches, such as the
formalism based on iterative solutions of time-dependent
states\cite{entin02} and the Keldysh formulation.\cite{vavilov01} Both
approaches can be used beyond the adiabatic approximation.

Experimentally, the first implementation of an electron pump was due
to Pothier {\it et al}. when charge was quantized due to Coulomb
blockade (CB) effects.\cite{pothier92} Adiabatic phase-coherent charge
pumping, though not quantized, was observed in open semiconductor
quantum dots \cite{switkes99} and in carbon nanotube quantum
dots.\cite{Leek05,Buitelaar08} Quantized charge pumping was recently
observed in AlGaAs/GaAs nanowires using a single-parameter
modulation,\cite{kaestner08} a result with potential applications to
metrology. An experimental realization of a quantum spin pump has also
been implemented.\cite{watson03}

Pumping through interacting systems, where the scattering approach
does not apply, has been much less studied so far. Using the
slave-boson mean-field approximation, Aono investigated the
spin-charge separation of adiabatic currents in the Kondo
regime.\cite{aono04} The behavior of the pumping current through a
quantum dot in the Kondo regime was studied both for
adiabatic\cite{Schiller08} and nonadiabatic systems\cite{Arrachea08}
using the Keldysh formalism. Quantum pumping was investigated both in
the CB regime \cite{brouwer05,cota05} as well as for almost open
quantum dots. \cite{aleiner98} The nonequilibrium Green's functions
technique has been employed to investigate adiabatic pumping through
interacting quantum dots in infinite $U$
systems.\cite{Splettstoesser05,sela06} The role of the Coulomb
interaction in the adiabatic pumping current has also been
investigated in the limit of weak tunneling and infinite-$U$ using
diagrammatic techniques.\cite{Splettstoesser06} The presence of
electron-electron interactions was shown to improve charge
quantization in one-dimensional disordered wires under certain
circumstances.\cite{Devillard08} The effects of the coupling of the
quantum dot to bosonic environments and its implications to charge
quantization were analyzed in Ref. \onlinecite{Fioretto08}. The
interplay of nonadiabacity and interaction effects on the pumping
current were also recently reported.\cite{Braun08,Cavalieri09}

In the present paper we investigate adiabatic charge pumping through
interacting quantum dots in the CB regime for temperatures much higher
than the Kondo temperature. We consider quantum dots with a single
level subjected to a finite Coulomb repulsion $U$ in the case of
double occupancy. We investigate the time dependence of the pumping
current by keeping $U$ finite, a scenario out of the domain of
validity of the theory developed in
Refs. \onlinecite{Splettstoesser05} and \onlinecite{sela06}. This
allows us to identify the relevant time scales controlling the current
amplitude in realistic situations. We develop a general formalism,
based on non-equilibrium Green's functions, to investigate the
influence of the Coulomb interaction on the adiabatic pumping
current. We discuss some applications and consequences of this
formulation and evaluate several quantities of interest numerically
for a range of parameters. Finally, the possibility of charge
quantization in the presence of a finite Coulomb repulsion is investigated. 
The study of charge quantization in the adiabatic regime is
interesting by its own, and is also a necessary step towards the
understanding of recent experiments\cite{kaestner08}
dealing with non-adiabatic pumping.

This paper is organized as follows. In Section \ref{ham} we present
the model used to calculate the time-dependent current flowing through
the quantum dot. Section \ref{sec:Gss} is devoted to the explicit
calculation of the relevant Green's functions. In Section
\ref{sec:adiabatic}, we apply this calculation to derive an expression
for the pumping current in the adiabatic approximation for systems
with finite $U$. The numerical evaluation of the current as well as a
discussion of its consequences and implications is presented in
Section \ref{sec:results}. Finally, Section \ref{sec:conclusions} is
devoted to a brief summary of our findings and concluding remarks.

\section{Model for transport in quantum dots}
\label{ham}

\begin{figure}[t]
\begin{center}
\includegraphics[width=8cm,angle=0]{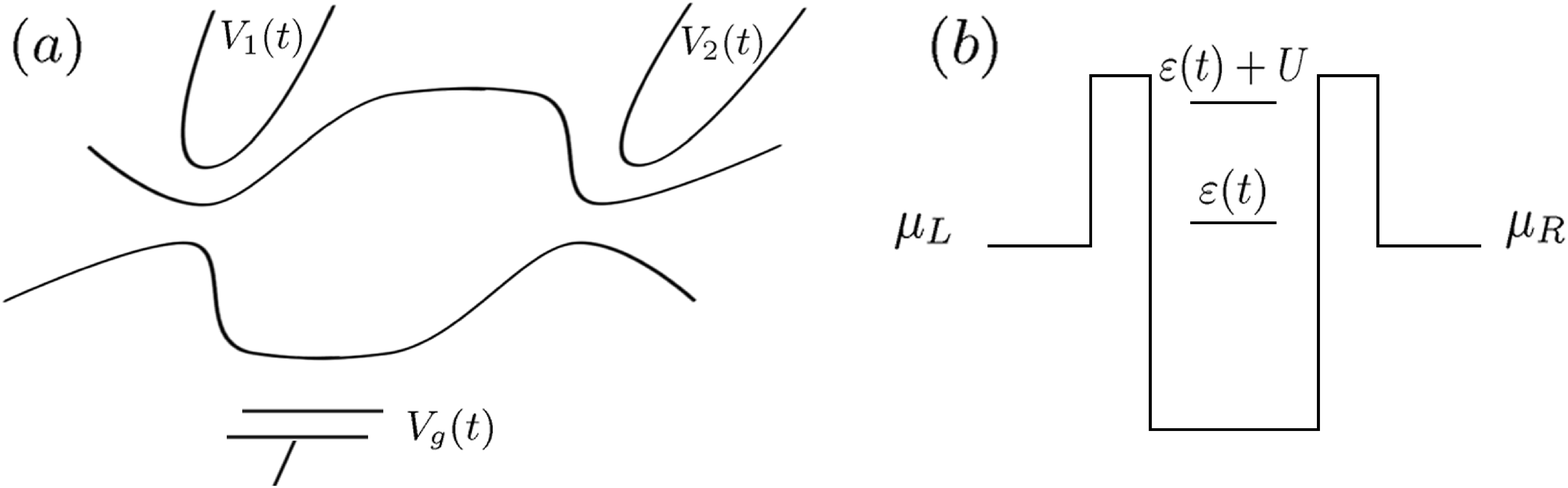}
\end{center} 
\caption{(a) Schematic view of a two-contact quantum dot coupled to a
time-dependent gate. (b) Sketch of the energy levels of the model
described in the text.}
\label{fig:1}
\end{figure}

We consider a quantum dot (QD) with a single, isolated resonance in
the Coulomb blockade regime, as schematically depicted in
Fig.~\ref{fig:1}. The potential in the dot is controlled by a
time-dependent gate voltage $V_g(t)$ such that the QD Hamiltonian
reads
\be \label{eq:HQD} H_{\rm dot} = \sum_{s=\uparrow,\downarrow}
\ve_s(t)\, d^\dagger_s d^{}_s + U\, n_\uparrow n_\downarrow, \ee
where $n_s = d^\dagger_s d_s$ is the number operator and
$d^\dagger_{s}$ ($d^{}_{s}$) is the creation (annihilation) operator
for an electron with energy $\ve_s(t)= \ve_{0s} - \eta\, e V_g(t)$ and
spin $s$ in the QD. Here, $e$ denotes the electron charge and $\eta$
is a lever arm factor for the gate voltage. Two single-channel leads
are attached to the QD. It is assumed that electrons in the leads are
noninteracting and obey the Hamiltonian
\be \label{eq:H_L} H_{\rm lead} = \sum_k \sum_{\alpha = {\rm L,R}}
\sum_{s=\uparrow,\downarrow} \ve_{k \alpha s}\, c^\dagger_{k
\alpha s} c^{}_{k \alpha s} \, , \ee
where $c^\dagger_{k\alpha s}$ and $c^{}_{k\alpha s}$ are, respectively
the creation and annihilation operators for electrons with momentum
$k$ and spin $s$ in the lead $\alpha$. The QD is separated from the
leads by tunneling barriers controlled by the lateral gates $V_1$ and
$V_2$ (see Fig.~\ref{fig:1}). The coupling Hamiltonian reads
\be \label{eq:H_LQD} H_{\rm lead-dot} = \sum_{k,\alpha,s} \left[
V_{k \alpha}(t)\, c_{k\alpha s}^{\dagger} d_s^{} + \rm
{H.c.}\right].  \ee
The tunneling matrix elements $V_{k\alpha}$ connect states in the
leads to the resonant state in the dot and are assumed to be spin
independent. The total Hamiltonian of our model is the sum of these
three contributions,
\be \label{eq:Htot} {\cal H} = H_{\rm lead} + H_{\rm dot}+ H_{\rm
lead-dot}.  \ee

The coupling between the states in the leads and those in the dot,
combined with the dot charging energy, turns the time evolution of the
system into a nontrivial many-body problem. As a result, we cannot
apply a single-particle formalism to describe the transport through
the system and the usual scattering-matrix formulation for pumping
currents~\cite{brouwer98} is inappropriate. To circumvent these
difficulties, we employ the Schwinger-Keldysh formalism and the
equation-of-motion method~\cite{ramm} to calculate the current through
an interacting quantum dot in the CB regime.

Our starting point is the general expression for the time-dependent
current in terms of the quantum dot Green's function $G_{s,s} (t,
t^\prime)$:\cite{HaugJauho96, Jauho94}
\begin{eqnarray}
\label{eq:current} J_\alpha(t) &\!\!= &\!\! -\frac{2e}{\hbar} \ {\rm
Im} \Big\{ \ \sum_{k, s} \int_{-\infty}^t dt^\prime\, V^\ast_{k
  \alpha}(t^\prime) e^{i\ve_{k\alpha s}(t-t^\prime)/\hbar} V_{k
  \alpha}(t) \nonumber \\ & & \times \left[ f_\alpha(\ve_{k\alpha
    s})\, G^r_{s,s}(t,t^\prime) + G^<_{s,s}(t, t^\prime) \right]
\Big\},
\end{eqnarray}
where $f_\alpha(E) = [e^{(E - \mu_\alpha)/k_BT} + 1]^{-1}$ is the
Fermi function for the lead $\alpha$ maintained at a chemical
potential $\mu_{\alpha}$ and temperature $T$ and $k_B$ is the 
Boltzmann constant. Throughout the text we consider pumping in
the absence of an external bias, that is,
$\mu_R=\mu_L=\varepsilon_F$. For convenience, we set
$\varepsilon_F=0$. The lesser, retarded, and advanced dot Green's
functions are defined as~\cite{ramm}
\ba 
\label{11} 
G^{<}_{s,s}(t, t^\prime) &\equiv & \frac{i}{\hbar} \left\langle
d^\dagger_{s}(t^\prime)\, d^{}_{s}(t) \right\rangle,
\nonumber \\
G^{r}_{s,s}(t, t^\prime) &\equiv & -\frac{i}{\hbar} \theta (t - t^\prime) \left\langle 
\{d^{}_{ s}(t),d^{\dagger}_{s} (t^\prime)\}\right\rangle, 
\nonumber \\ 
G^{a}_{s,s}(t, t^\prime) &\equiv & \frac{i}{\hbar}\theta (t^\prime - t) \left\langle 
\{d^{}_{s}(t),d^{\dagger}_{s} (t^\prime)\} \right\rangle.  \ea
Now it remains to compute the Green's function $G_{s,s}(t, t^\prime)$
which involves the quantum dot states. This is where the many-body
aspects of the problem make their way into the pumping
current. Section \ref{sec:Gss} is devoted to this issue.

\section{Calculation of $G_{s, s}$} 
\label{sec:Gss}

The current in Eq.~(\ref{eq:current}) is given in terms of the quantum
dot Green's functions $G^{r}_{s,s}(t, t^{\prime})$ and $G^{<}_{s,s}(t,
t^{\prime})$. To write expressions for them, we start by calculating
the time-ordered Green's function $G_{s,s} (t, t^{\prime})$, defined
as \cite{ramm}
\be 
\label{eq:defGss} 
G_{s,s}(t, t^\prime) \equiv -\frac{i}{\hbar} \left \langle {\cal T}\!
\left[ d^{}_{s}(t)\, d^\dagger_{s}(t^\prime) \right] \right\rangle,
\ee
where ${\cal T}$ is the time-ordering operator. The equation-of-motion
for $G_{s,s}$ is
\begin{eqnarray} 
\label{eq:EOM_Gss} 
\!\!\!\left[ i\hbar \frac{\partial}{\partial t} - \ve_s(t) \right]
G_{s,s}(t,t^\prime) & = & \delta(t-t^\prime) + U\,
G^{(2)}_{s\bar{s},s}(t,t^\prime) \nonumber \\ & +&  \sum_{k, \alpha}
V_{k\alpha}^\ast(t)\, G_{k \alpha s, s}(t, t^\prime).
\end{eqnarray}
In Eq.~(\ref{eq:EOM_Gss}) we have introduced the ``contact"
time-ordered Green's function 
\be
\label{12a} 
G_{s, k \alpha s}(t, t^\prime) \equiv -\frac{i}{\hbar} \left\langle
{\cal T} \left[ d^{}_{s}(t)\, c^{\dagger}_{k \alpha s}(t^\prime)
  \right] \right\rangle, \ee
which obeys the equation-of-motion
\be \label{eq:EOM_G0k} \left(-i \hbar \frac{\partial}{\partial
t^\prime} - \ve_{k\alpha s}\right) G_{s, k \alpha s}(t, t^\prime)
= V_{k \alpha}^\ast(t^\prime)\, G_{s,s}(t, t^\prime), \ee
as well as the second-order correlation function
\be \label{eq:defG2} G^{(2)}_{s\bar{s},s}(t, t^\prime)\equiv
-\frac{i}{\hbar} \left\langle {\cal T}\!\left[ d^{}_s(t)\,
n^{}_{\bar{s}}(t)\, d^{\dagger}_{s}(t^\prime)\right]\right\rangle,
\ee
that involves four fermionic operators and is generated by the
interaction term $Un_{\downarrow}n_{\uparrow}$. The same interaction
term leads to the appearance of even higher order correlation
functions in the equation-of-motion for $G^{(2)}$, namely,
\begin{widetext}
\begin{equation} 
\left[ i\hbar \frac{\partial}{\partial t} - \ve_s(t) - U\right]
G^{(2)}_{s\bar{s},s}(t,t^\prime) =  \delta(t-t^\prime)\langle
n_{\bar{s}}(t)\rangle + \sum_{k \alpha} \left[ V_{k \alpha}^\ast
\Gamma_{1;k\alpha s}^{(2)}(t,t^\prime) \right] + \sum_{k \alpha}
\left[ V_{k \alpha} \Gamma_{2;k\alpha s}^{(2)}(t,t^\prime)
- V_{k \alpha}^\ast \Gamma_{3;k\alpha
s}^{(2)}(t,t^\prime) \right],
\end{equation}
\end{widetext}
where the occupation number is defined as
\be \langle n_s(t) \rangle = \langle d_s^\dagger(t) d_s(t) \rangle
\equiv i\hbar\, G_{s,s}^<(t,t) \ee
and we have introduced three lead-dot correlation
functions,\cite{HaugJauho96}
\begin{equation} 
\Gamma^{(2)}_{1;k \alpha s}(t,t^\prime) \equiv -\frac{i}{\hbar}
\left\langle {\cal T}\! \left[c^{}_{k \alpha s}(t)n^{}_{\bar{s}}(t)
  d^{\dagger}_{s}(t^\prime) \right]\right\rangle,
\end{equation}
\begin{equation}
\Gamma^{(2)}_{2;k \alpha s}(t,t^\prime) \equiv -\frac{i}{\hbar}
\left\langle {\cal T}\! \left[c^{\dagger}_{k \alpha
    \bar{s}}(t)d^{}_{s}(t) d^{}_{\bar{s}}(t) d^{\dagger}_{s}(t^\prime)
  \right]\right\rangle,
\end{equation}
and
\begin{equation}
\Gamma^{(2)}_{3;k \alpha s}(t,t^\prime) \equiv -\frac{i}{\hbar}
  \left\langle {\cal T}\!  \left[c^{}_{k \alpha
  \bar{s}}(t)d^{\dagger}_{\bar{s}}(t) d^{}_{s}(t)
  d^{\dagger}_{s}(t^\prime) \right]\right\rangle.
\end{equation}
At this level, one can verify that the equations-of-motion do not
close. Going to the next level, one obtains new (higher order)
correlation functions and even more complicated expressions. To solve
this problem, we shall recur to an approximate scheme, namely the
mean-field approximation.

\subsection{Formal solution of the equations-of-motion within the Hartree 
approximation}
\label{sec:HF}

We now focus on the Coulomb blockade regime and neglect spin
correlations in the leads. That is, we assume that the Kondo
temperature,\cite{Hewson} $T_K \sim U \sqrt{\Gamma/2U}\exp
\big(-\pi|\varepsilon_s| (\varepsilon_s + U)/2U\Gamma\big)$ is very
low, $T_K \ll T$. As usual, $\Gamma$ stands for the quantum dots
resonance linewidth which will be precisely defined in
Sec.~\ref{sec:adiabatic}. Hence, with respect to Kondo correlations,
we are in the high-temperature regime and the mean-field approximation
is expected to be valid. Within this approximation, one can write the
$\Gamma^{(2)}$'s as
\begin{eqnarray}
\label{eq:mean_field} 
\Gamma^{(2){\rm mf}}_{1;k \alpha s} (t,t^\prime)= \langle n_{\bar{s}} (t)
\rangle \,G_{k \alpha s, s}(t,t^\prime)
\end{eqnarray}
and
\begin{eqnarray}
\Gamma^{(2){\rm mf}}_{2;k \alpha s} (t,t^\prime)= \Gamma^{(2){\rm
mf}}_{3;k \alpha s} (t,t^\prime)= 0.
\end{eqnarray}
It has been shown that Kondo correlations are still absent in the next
order of the equations-of-motion hierarchical
truncation.\cite{Meir91,HaugJauho96} The latter dresses the Green's
functions self-energies with higher order terms in $V$ that include,
for instance, cotunneling processes. As long as $\varepsilon_s$ is of
the order of $k_BT$, we have verified that these contributions give
only small corrections to the Hartree mean-field
approximation.\cite{HaugJauho96} Thus, we write
\ba \label{eq:EOM_G2} 
&&\!\!\!\!\!\!\!\!\!\!\!\!
\left[ i\hbar \frac{\partial}{\partial t}- \ve_s(t) - U \right] 
G^{(2){\rm mf}}_{s\bar{s},s}(t,t^\prime) = \nonumber\\&&
\langle n_{\bar{s}}(t)\rangle \left[ \delta(t-t^\prime)+ \sum_{k
    \alpha} V_{k \alpha}^*(t) \,G_{k \alpha s, s}(t,t^\prime)\right],
\ea
where the occupation number $\langle n_{\bar{s}} (t)\rangle$ has to be
determined self-consistently for all times. Equations
(\ref{eq:EOM_Gss}), (\ref{eq:EOM_G0k}), and (\ref{eq:EOM_G2}) form a
closed set of equations-of-motion that determines the time-ordered
Green's function $G_{s,s}$. Using analytical continuation and the
Langreth rules~\cite{Langreth66,HaugJauho96} we can then find the
Green's functions $G^r_{s,s}$ and $G^<_{s,s}$ that appear in the
expressions for the current, Eq.~(\ref{eq:current}). For convenience,
let us define two auxiliary time-ordered Green's functions $g_{s}$ and
$g_{s}^U$ that obey the equations-of-motions
\be 
\label{eq:defsmallg0} 
\left[ i\hbar \frac{\partial}{\partial t} - \ve_s(t) \right] g_{s}(t,
t^\prime) = \delta(t- t^\prime) \ee
and
\be \label{eq:defsmallg0U} \left[ i\hbar \frac{\partial}{\partial
t} - \ve_{s}(t) - U \right] g_{s}^U(t, t^\prime) = \delta(t -
t^\prime), \ee
respectively. By analytical continuation into the complex plane, we
can rewrite Eq.~(\ref{eq:EOM_G2}) as
\begin{eqnarray}
&&\!\!\!\!
G_{s\bar{s},s}^{(2)\rm{mf}}(\tau, \tau^\prime) = g_{s}^U(\tau,
  \tau^\prime) \langle n_{\bar{s}}(\tau^\prime)\rangle \\ &&
  +\sum_{k,\alpha} \int d\tau_1\, g_{s}^U (\tau,\tau_1)\, \langle
  n_{\bar{s}}(\tau_1) \rangle\, V_{k\alpha}^\ast(\tau_1)\, G_{k\alpha
    s,s}(\tau_1,\tau^\prime).  \nonumber
\end{eqnarray}
The equation for $G_{k\alpha s,s}(\tau_1,\tau^\prime)$ can also be
obtained in a similar manner. Using Eq. (\ref{eq:EOM_G0k}), the
equation-of-motion for the time-ordered Green's function for free
electrons in the leads, namely,
\be \label{eq:EOM_g} \left(-i \hbar \frac{\partial}{\partial
t^\prime} - \ve_{k\alpha s} \right) g_{k \alpha s}(t,t^\prime) =
\delta(t - t^\prime), \ee
and the rules of analytical continuation, we conclude that the
contour-ordered Green's function $G_{s, k \alpha s}(\tau,
\tau^\prime)$ obeys the equation
\be \label{17a} G_{s, k \alpha s}(\tau,\tau^\prime) = \int d\tau_1\,
G_{ss}(\tau, \tau_1)\, V^{*}_{k \alpha}(\tau_1) \, g_{k \alpha
  s}(\tau_1, \tau^\prime), \ee
while its counterpart is given by
\be \label{17b} G_{k \alpha s,s}(\tau,\tau^\prime) = \int d\tau_1\,
g_{k \alpha s}(\tau, \tau_1)\, V_{k \alpha}(\tau_1)\, G_{s,s}(\tau_1,
\tau^\prime). \ee
In all these cases the integration paths run over the Keldysh contour
discussed in Refs.~\onlinecite{HaugJauho96} and
\onlinecite{Hernandez07}.

Now the equations-of-motions close since both $G^{(2)\rm{mf}}$ and
$G_{k \alpha s,s}$ are expressed in terms of $G_{s,s}$ and free
Green's functions. By introducing the renormalized single-electron
resolvent
\be \label{eq:g0sbarr} \bar{g}_{s}(\tau, \tau^\prime) \equiv
g_{s}(\tau, \tau^\prime) + U \left\langle n_{\bar{s}}(\tau^\prime)
\right\rangle \int d\tau_1 \, g_{s}(\tau, \tau_1)\, g_{s}^U(\tau_1,
\tau^\prime), \ee
we write, after a little algebra, a Dyson-like equation for $G_{s,s}$,
\begin{eqnarray} 
\label{eq:Gbigss} 
G_{s,s}(\tau, \tau^\prime) & = & \bar{g}_{s}(\tau, \tau^\prime) + \int
d\tau_1 \int d\tau_2\, \bar{g}_{s}(\tau, \tau_1) \nonumber \\ & &
\times\ \Sigma_{ss}(\tau_1, \tau_2) G_{s,s}(\tau_2, \tau^\prime),
\end{eqnarray}
with the self-energy defined as
\be \label{eq:selfenergyfinal} \Sigma_{ss}(\tau, \tau^\prime) =
\sum_{k\alpha} V_{k\alpha}^\ast(\tau\,) g_{k\alpha s}(\tau
-\tau^\prime)\, V_{k\alpha}(\tau^\prime).  \ee
The rather peculiar structure of our solution is noteworthy. The
auxiliary Green's function $\bar{g}_{s}$, Eq.~(\ref{eq:g0sbarr}), is
not a free propagator since it contains a term involving $\langle
n_{\bar{s}} \rangle$ that arises from the mean-field approximation and
has to be calculated self consistently. The self energy carries
information about the coupling to the leads and can be calculated
independently of the state of the dot. Hence it does not contain
information about the many-body character of the problem.

In Section \ref{sec:adiabatic}, we shall specialize the calculation to
the adiabatic regime, first by explicitly obtaining an expression for
the Green's functions involved in Eqs. (\ref{eq:g0sbarr}) and
(\ref{eq:Gbigss}) and then by evaluating the current,
Eq.~(\ref{eq:current}).

\section{Electronic transport in the adiabatic approximation}
\label{sec:adiabatic}

The two important time scales in the problem of charge pumping through
non-interacting quantum dots are the mean dwell time of an electron
inside the dot (lifetime of the resonant state), $\tau_{D}$, and the
inverse of the characteristic pumping frequency, $\tau_{\rm pump} =
2\pi/\omega_{\rm pump}$. In typical experimental setups, the pumping
frequency $\omega_{\rm pump}$ lies in the range between 10 MHz to 1
GHz.\cite{switkes99} For $\omega_{\rm pump}/2\pi = 100$ MHz, one has
$\tau_{\rm pump} = 10$ ns. The mean dwell time is given by the inverse
of the resonance width $\Gamma$. To estimate it, let us first recall
that the dot single-particle mean level spacing is $\Delta =
2\pi\hbar^2/({\mathcal A}m^*)$, where ${\mathcal A}$ is the dot
effective area and $m^* = 0.067 m_e$ for GaAs. We obtain $\Delta
\approx 7.6\, \mu{\rm eV}(\mu{\rm m})^2/{\mathcal A}$, where
${\mathcal A}$ is given in square microns. For the Coulomb blockade
regime, typical resonance widths are $\Gamma = 0.01 - 0.1 \Delta$. As
a result, $\tau_D = \hbar/\Gamma \approx 0.8 - 8$ ns$(\mu{\rm
  m})^2/{\mathcal A}$ for most devices. For ${\mathcal A}$ much
smaller than 1 $(\mu$m$)^2$, we find that $\tau_{\rm pump} \gg
\tau_{D}$. In this case we can safely employ the so-called
\emph{adiabatic approximation}, which precisely relies on the fact
that the time scale over which the system parameters vary is large
compared to the lifetime of the electron in the dot.

\subsection{Adiabatic approximation for the Green's functions}
\label{sec:meantime}

A convenient way to separate slow and fast times scales is to
reparametrize the Green's functions as
\be \label{eq:meantime} G(t,t')\rightarrow G \left( t-t^\prime
,\frac{t + t^\prime}{2} \right), \ee
that is, the time variables are replaced by a (fast) time difference
$\delta t = t-t^\prime$ and a slow mean time $\bar{t}=(t+t')/2$. We
implement the adiabatic approximation to lowest order by expanding the
Green's functions up to linear order in the slow variables, namely,
\begin{eqnarray}
\label{eq:Gexpansion} 
G\left(t - t^{\prime}, \frac{t + t^{\prime}}{2}\right) & \approx &
  G\left(t- t^{\prime}, t\right) \nonumber \\ & & + \left(
  \frac{t^{\prime} - t}{2} \right) \frac{\partial G}{\partial \bar{t}}
  \left(t - t^{\prime},\bar{t}\right) \Big|_{\bar{t} = t}.
\end{eqnarray}
In what follows we formally write 
\begin{equation}
G(t-t^\prime, \bar{t}) = G^{(0)}(t-t^\prime, \bar{t}) +
G^{(1)}(t-t^\prime, \bar{t}),
\end{equation}
where the zeroth order refers to equilibrium quantities, while the
adiabatic contributions, linear in the slow time variable (and in our
case proportional to the pumping frequency), are collected in the
first-order correction. The accuracy of our approximation can be
tested by inspecting higher-order terms. We will return to this issue
in Sec.~\ref{sec:results}, when we present our results.

Let us now describe how the approximate scheme works. Using the
mean-time parametrization, we write Eq.~(\ref{eq:g0sbarr}) as
\begin{widetext}
\begin{equation}
 \bar{g}_{s} \left( t - t^\prime , \bar{t}\, \right) = g_{s} \left(t -
 t^\prime, \bar{t}\, \right) + U \left\langle
 n_{\bar{s}}(\bar{t})\right\rangle \int_{-\infty}^\infty dt_1 \, g_{s}
 \left( t - t_1, \frac{t + t_1}{2} \right)\, g_{s}^U \left( t_1 -
 t^\prime , \frac{t_1 + t^\prime}{2} \right).
\end{equation}
Expanding $\bar{g}_{s}$ in the slow variables as in
Eq.~(\ref{eq:Gexpansion}) and taking the Fourier transform with
respect to the fast variable, namely, $g(\omega, \bar{t}) =
\int_{-\infty}^\infty d(t-t^{\prime})\, g(t-t^{\prime},\bar{t})
\exp[i\omega(t-t^{\prime})]$, we obtain
\be 
\bar{g}_{s} (\omega, \bar{t} ) = \bar{g}^{(0)}_{s} (\omega, \bar{t}) +
\bar{g}^{(1)}_{s} (\omega, \bar{t}), \ee
with
\be 
\label{go57} 
\bar{g}^{(0)}_{s}(\omega, \bar{t}\, ) = g_{s}^{(0)}(\omega, \bar{t}\,
) + U \left\langle n_{\bar{s}}^{(0)}(\bar{t}) \right\rangle \,
g_{s}^{(0)}(\omega ,\bar{t}\, )\, g_{s}^{U(0)}(\omega , \bar{t} \,)
\ee
and
\ba 
\label{go58}
\bar{g}^{(1)}_{s} (\omega, \bar{t} ) & = & g_{s}^{(1)} (\omega,
\bar{t} ) + U \left[ \left\langle n_{\bar{s}}^{(1)}(\bar{t})
\right\rangle \, g_{s}^{(0)}\, g_{s}^{U(0)} +\, \langle
n_{\bar{s}}^{(0)}(\bar{t})\rangle \, g_{s}^{(1)}\, g_{s}^{U(0)} +
\left\langle n_{\bar{s}}^{(0)}(\bar{t}) \right\rangle \, g_{s}^{(0)}\,
g_{s}^{U(1)} \right] \\ & & +\ \frac{i\hbar}{2} U \left[\frac{\partial
\left\langle n_{\bar{s}}(\bar{t}) \right\rangle}{\partial \bar{t}}
\frac{\partial \left( g_s g_s^U \right) }{\partial \omega}-
\left\langle n_{\bar{s}}^{(0)}(\bar{t}) \right\rangle \,
\frac{\partial g_{s}^{(0)}}{\partial \bar{t}}\, \frac{\partial
g_{s}^{U(0)}}{\partial \omega} + \left\langle
n_{\bar{s}}^{(0)}(\bar{t}) \right\rangle \, \frac{\partial
g_{s}^{(0)}}{\partial \omega} \frac{\partial g_{s}^{U(0)}} {\partial
\bar{t}} \right], \nonumber \ea
\end{widetext}
where 
\be \left\langle n_{\bar{s}}(\bar{t}) \right\rangle= \left\langle
n_{\bar{s}}^{(0)}(\bar{t}) \right\rangle + \left\langle
n_{\bar{s}}^{(1)}(\bar{t}) \right\rangle \ee
is introduced following the same principle as the one described after
Eq.~(\ref{eq:Gexpansion}).

Equation (\ref{go58}) is further simplified by the fact that the
lowest order corrections to terms involving $g_{s}^{(1)}$ and
$g_{s}^{U(1)}$ vanish for the retarded component. To demonstrate this,
let us consider the retarded component
\be 
g^{r}_{0,{s}}(t-t')=-\frac{i}{\hbar}\Theta(t-t')
\exp\left[-\frac{i}{\hbar}\int_{t'}^t dt_1\epsilon_s(t_1)\right].
\ee
Expanding $\epsilon_s(t_1)$ around the mean-time $\bar{t}=
(t+t^\prime)/2$, namely, $\epsilon_s(t_1)=\epsilon_s (\bar{t}) +
\dot\epsilon_s (\bar{t}) (t_1- \bar{t}) $ we obtain 
\be
\int^{t}_{t^{\prime}}dt_1\epsilon_s(t_1)= \epsilon_s(\bar{t})\bar{t} +
O (\ddot{\epsilon}),
\ee 
so that $g_{s}^{(1)r(a)} = g_{s}^{U(1)r(a)} =0$.\cite{footnote1} This
simplification shows the advantage of the mean-time parametrization,
Eq.~(\ref{eq:meantime}), with respect to other parameterizations, such
as the one chosen in Ref.~\onlinecite{Splettstoesser05}.

After these simplifications, we obtain for the advanced and retarded
components
\begin{eqnarray}
\bar{g}^{(0)r(a)}_{s}(\omega, \bar{t} ) & = & g_{s}^{(0)r(a)}(\omega,
\bar{t} ) + \left\langle n_{\bar{s}}^{(0)}(\bar{t}) \right\rangle
\nonumber \\ & & \times\ g_{s}^{(0)r(a)}(\omega ,\bar{t}) U
g_{s}^{U(0)r(a)}(\omega , \bar{t})
\end{eqnarray}
and 
\ba \bar{g}^{(1)r(a)}_{s}(\omega, \bar{t}) &=& \left\langle
n_{\bar{s}}^{(1)}(\bar{t}) \right\rangle \, g_{s}^{(0)r(a)}(\omega
,\bar{t}) U g_{s}^{U(0)r(a)}(\omega , \bar{t}) \nonumber \\ & & +\
\frac{i\hbar}{2}U \frac{\partial \left\langle
n_{\bar{s}}^{(0)}(\bar{t}) \right\rangle} {\partial \bar{t}}
\frac{\partial}{\partial \omega} \left[ g_s^{(0)\, r(a)} (\omega ,
\bar{t}) \right. \nonumber \\ & & \left. \times\ g_s^{U\, (0)\, r(a)}
(\omega , \bar{t})\right].  \ea
For the lesser components, we employ the fluctuation-dissipation
theorem to write 
\ba \label{g1lesseraux0} \bar{g}^{(0)<}_{s}(\omega, \bar{t} ) &=&
f(\omega)\left[ \bar{g}_{s}^{(0)a}(\omega ,\bar{t}) -
  \bar{g}_{s}^{(0)r}(\omega ,\bar{t})\right], \ea 
and apply the Langreth rules to Eq.~(\ref{go58}) to obtain
\begin{widetext}
\ba \bar{g}^{(1)<}_{s}(\omega, \bar{t}) &=& U \langle
n_{\bar{s}}^{(1)}(\bar{t})\rangle f(\omega) \left[ g_{s}^{(0)a}(\omega
,\bar{t}) g_{s}^{U(0)a}(\omega , \bar{t}) - g_{s}^{(0)r}(\omega
,\bar{t}) g_{s}^{U(0)r}(\omega ,\bar{t}) \right] \nonumber \\ & & +\
\frac{i\hbar}{2} U \frac{\partial \langle
n_{\bar{s}}^{(0)}(\bar{t})\rangle} {\partial \bar{t}}
\frac{\partial}{\partial \omega} \left\{ f(\omega)\left[ g_s^{(0)\, a}
(\omega , \bar{t}) g_s^{U\, (0)\, a} (\omega , \bar{t})- g_s^{(0)\,
r}(\omega , \bar{t}) g_s^{U\, (0)\, r} (\omega , \bar{t})
\right]\right\} \nonumber \\ & & +\ \frac{i\hbar}{2}U\langle
n_{\bar{s}}^{(0)}(\bar{t})\rangle \frac{\partial f(\omega)}{\partial
\omega}\left\{ \left[g_{s}^{(0)a}(\omega ,\bar{t}) -
g_{s}^{(0)r}(\omega ,\bar{t})\right] \frac{\partial
g_{s}^{U(0)a}}{\partial \bar{t}}(\omega,\bar{t}) - \frac{\partial
g_{s}^{(0)r}}{\partial \bar{t}}(\omega ,\bar{t}) \left[
g_{s}^{U(0)a}(\omega ,\bar{t}) \right. \right. \nonumber \\ & &
\left. \left. -\ g_{s}^{U(0)r}(\omega ,\bar{t}) \right] \right\} .
 \label{g1lesseraux}
\ea
Here $f(\omega)= [\exp(\hbar \omega/k_BT) +1]^{-1}$.

We proceed in the same way to obtain an expression for $G_{s,s}$. The
result is
\be \label{Gtotal} G_{s,s}(\omega, \bar{t})= G^{(0)}_{s,s}(\omega,
\bar{t})+G^{(1)}_{s,s}(\omega, \bar{t}), \ee
with
\be \label{G0final} G^{(0)}_{s,s}(\omega, \bar{t}) =
\bar{g}^{(0)}_{s}(\omega , \bar{t}) + \bar{g}^{(0)}_{s}(\omega ,
\bar{t}) \Sigma^{(0)}_{s,s}(\omega , \bar{t})
G^{(0)}_{s,s}(\omega, \bar{t})
\ee
and
\ba 
\label{G1} 
G^{(1)}_{s,s}(\omega,\bar{t}) & = & \bar{g}^{(1)}_{s}(\omega, \bar{t})
+ \bar{g}^{(1)}_{s}(\omega,\bar{t})
\Sigma_s(\omega,\bar{t})G^{(0)}_{ss}(\omega,\bar{t}) +
\bar{g}^{(0)}_{s}(\omega,\bar{t})
\Sigma_s(\omega,\bar{t})G^{(1)}_{ss}(\omega,\bar{t}) \nonumber \\ & &
-\ \frac{i\hbar}{2}\,\,\frac{\partial \bar{g}^{(0)}_{s}}{\partial
\bar{t}}
(\omega,\bar{t})\,\frac{\partial}{\partial\omega}\!\left[\Sigma_s(\omega,
\bar{t})G^{(0)}_{ss}(\omega,\bar{t})\right] +
\frac{i\hbar}{2}\,\,\frac{\partial}{\partial
\omega}\left[\bar{g}^{(0)}_{s}
(\omega,\bar{t})\Sigma_s(\omega,\bar{t})\right] \frac{\partial
G^{(0)}_{ss}}{\partial \bar{t}} (\omega,\bar{t}) \nonumber \\ & & +\
\frac{i\hbar}{2}\frac{\partial \bar{g}^{(0)}_{s}}{\partial
\omega}(\omega , \bar{t}) \frac{\partial \Sigma_s}{\partial
\bar{t}}(\omega,\bar{t}) G^{(0)}_{ss}(\omega,\bar{t}) -
\frac{i\hbar}{2}\bar{g}^{(0)}_{s}(\omega , \bar{t})\frac{\partial
\Sigma_s}{\partial \bar{t}} (\omega,\bar{t}) \frac{\partial
G^{(0)}_{ss}}{\partial \omega}(\omega,\bar{t}) \nonumber \\ & & -\
\frac{i\hbar}{2}\bar{g}^0_{s}(\omega , \bar{t}) {\cal S}^{(1)}
(\omega, \bar{t}) G^{(0)}_{s,s}(\omega,\bar{t}).  \ea
\end{widetext}
In Eq.~(\ref{G1}) we have introduced 
\be {\cal S}^{(1)} (\omega,
\bar{t}) = \sum_{k\alpha}
\left[\dot{V}^*_{k\alpha}(\bar{t}){V}_{k\alpha}(\bar{t}) -
{\rm H.c.}  \right]
\frac{\partial g_{k\alpha s}}{\partial \omega}(\omega, \bar{t}).
\ee
In what follows we use the wide-band approximation, where
$\Sigma(\omega,t) \rightarrow \Sigma(t)$, in which case the above
equations are simplified further.

From Eqs.~(\ref{G0final}) and (\ref{G1}), we obtain $G^r$ and $G^<$,
which are needed to calculate $J_\alpha$, Eq.~(\ref{eq:current}), in
the adiabatic approximation for the Coulomb blockade regime. Since the
zeroth order terms are essentially equilibrium quantities, we are
allowed to use the fluctuation-dissipation theorem to compute
$G^{(0)<}$ without much effort: $G^{(0)<}(\omega, t) = -2i f(\omega)
{\rm Im} \left[ G^{(0)r}(\omega, t) \right]$. For $G^{(1)<}$ this is
no longer possible and we have to use the Langreth rules. The
resulting expressions are rather long and will be omitted here.

The occupation numbers $\langle n^{(0)}_{\bar{s}}\rangle$ and $\langle
n^{(1)}_{\bar{s}}\rangle$ that appear in Eqs.~(\ref{G0final}) and
(\ref{G1}) are calculated self consistently using
\be 
\label{FD} 
\left\langle n^{(i)}_s (t) \right\rangle = \int_{-\infty}^\infty\frac{d\omega}{2
\pi i}G^{(i)<}_{s,s}(\omega, t), \ee
where $i=0$ or 1. In the absence of an external magnetic field, which
is the case considered here, $\left\langle n^{(i)}_{\bar{s}}
\right\rangle = \left\langle n^{(i)}_s \right\rangle$.

For later convenience, we assume the couplings $V_{k\alpha}$ to be
energy independent and use the flat and wide band approximation to
define
\be \Gamma_{\alpha}(\epsilon, t) = 2 \pi \left|V_{\alpha}
(\epsilon, t) \right|^{2} \rho_{\alpha} \cong 2 \pi \left|V_{\alpha}
  (t)\right|^{2} \rho_{\alpha} \equiv \Gamma_{\alpha}(t),
\ee
with $\rho_{\alpha}$ denoting the density of states in the lead
$\alpha$. We also introduce
\be 
\Gamma(t) = \sum_{\alpha} \Gamma_{\alpha} (t)
\label{couplings} 
\ee
as the total decay width. As we discuss next, the current in
Eq.~(\ref{eq:current}) is easily cast in terms of these quantities.

\subsection{Current in the adiabatic approximation}
\label{sec:adiabatic_current}

To evaluate the time integral in the general expression for the
current, we proceed as in Eq.~(\ref{eq:Gexpansion}) and expand all
terms in the integrand to linear order in the slow variables. The
resulting expression for the pumped current depends explicitly on
$G^<(\omega,t)$ and $G^r(\omega, t)$. Since $G^<$ is related to
occupations (and hence to fluctuations) and $G^r$ to dissipation, as
shown by standard linear response theory, it is natural to break the
current into two parts,
\be J_{\alpha} (t) \equiv J^{\rm fl}_{\alpha} (t) + J^{\rm dis}_{\alpha}(t),
\label{currentFD}\ee
where the fluctuation term is
\ba J^{\rm fl}_{\alpha } (t) & =& -\frac{2e}{\hbar}\sum_s \mbox{Im}\! \left[
  \frac{\Gamma_{\alpha}(t)}{2} \int_{-\infty}^{\infty} \frac{d
  \omega}{2 \pi} G_{s,s}^{<}(\omega,t) \right] \nonumber\\ &=&
  -\frac{e}{\hbar} \Gamma_{\alpha}(t)\sum_s \left\langle
  n_{s} (t)\right\rangle
\label{flucuation} 
\ea
while the dissipation term is given by
\begin{multline}
J^{\rm dis}_{\alpha }(t)  = -\frac{2e}{\hbar}\sum_s\mbox{Im}\Bigg\{
\int_{-\infty}^{\infty}\frac{d\omega}{2\pi} f(\omega) 
\Bigg[\Gamma_\alpha(t)G^r_{s,s}(\omega,t) \\ 
 +\ \frac{i\hbar}{2} \frac{d}{dt} \left(\Gamma_\alpha(t)
\frac{\partial G^r_{s,s}}{\partial \omega} (\omega,t) \right)
 \Bigg] \Bigg\} + O (\partial^2_\omega \partial^2_t). 
\label{dissipation}
\end{multline}
Now we are ready to use the adiabatic expansion for the Green's
function, $G_{s,s} = G_{s,s}^{(0)}+G_{s,s}^{(1)}$, and to identify the
zeroth and the first-order contributions to the pumped current,
$J^{(0)}$ and $J^{(1)}$, respectively. It can be shown that the zeroth
order current vanishes, as expected by the fluctuation-dissipation
theorem.

The first-order contribution to the current due to fluctuation is
given by
\begin{eqnarray} J^{(1){\rm fl}}_{\alpha} (t) 
&=& -\frac{e}{\hbar} \Gamma_{\alpha}(t)\sum_s 
  \left\langle n^{(1)}_{\bar{s}} (t) \right\rangle,
\label{fluctuation1} \end{eqnarray}
while the first-order dissipation term is given by
\be J^{(1){\rm dis}}_{\alpha} (t) = J^{(1a){\rm dis}}_{\alpha} (t)
+ J^{(1b){\rm dis}}_{\alpha} (t), \ee 
where
\be
J^{(1a){\rm dis}}_{\alpha } (t) = -\frac{2e}{\hbar} \sum_s\mbox{Im} \left[
  \Gamma_{\alpha} (t) \int_{-\infty}^{\infty} \frac{d \omega}{2 \pi}
  f(\omega) G_{s,s}^{(1)r}(\omega,t)
  \right] 
\label{dissipation1a}
\ee
and
\begin{eqnarray} \!\!\!
J^{(1b){\rm dis}}_{\alpha} (t) = -e \sum_s \mbox{Re}&&\!\!\!\!\!
\left\{ \int_{-\infty}^{\infty} \frac{d \omega}{2 \pi} \left(
-\frac{\partial f}{\partial \omega} \right) 
\right. \nonumber \\ & & \left. \!\!\!\!
\times\frac{d}{dt}\left[
\Gamma_{\alpha} (t) G_{s,s}^{(0) r}(\omega,t) \right]\right\} .
\label{dissipation1b}
\end{eqnarray}
The reason for breaking the dissipation term into two contributions is
that $J^{(1b){\rm dis}}_{\alpha} (t)$ is a total derivative in
time. Integrated over a pumping period, this current term does not
contribute to the pumped charge. This provides a good check for the
numerical calculations presented in Sec.~\ref{sec:results}. We also
successfully verified that our analytical expressions yield the same
results as other pumping formulations
\cite{brouwer98,Splettstoesser05} in the $U \rightarrow 0$ limit.

Equations (\ref{currentFD}), (\ref{fluctuation1}), and
(\ref{dissipation1a}) constitute the principal results of this
paper. In the following, we will use these expressions to investigate
the role of interactions on the pumped current. Specifically, we will
study how interactions affect the dependence of the pumped current on
$U$, temperature, and the phase difference between the pumping
perturbations.

\section{Results and discussions}
\label{sec:results}

In this Section we compute numerically the pumping current,
Eq.~(\ref{currentFD}), and investigate the dependence of the magnitude
of the leading contribution to the total charge pumped per cycle,
\be Q = \int_0^{\tau_{\rm pump}} dt\, J^{(1)}_L(t), \ee
on several model parameters. In particular, we discuss in which
conditions the pumped charge can be quantized to its maximum value,
$|e|$. To accomplish this goal, we consider the following
parametrization for tunnel couplings:
\be \Gamma_{\alpha} (t) = \Gamma_{0,\alpha} + \Delta \Gamma_{\alpha}
\cos\left(\Omega t + \phi_{\alpha}\right), \label{gamamodel}\ee
where $\alpha = R,L$ and $\Gamma_{0,\alpha}$ and $\Delta
\Gamma_{\alpha}$ are real constants. We also assume that the quantum
dot resonance energy varies in time as
\be \label{eq:e(t)}
 \varepsilon(t) = \varepsilon_0 + \varepsilon_{1}\cos(\Omega t).\ee 
Notice that since $\varepsilon_s=\varepsilon_{\bar{s}}$, we have
dropped the spin index. In the following, all parameters are chosen to
ensure that the system is clearly in Coulomb blockade regime, $\Gamma
\ll U$. Typically, we take $\Gamma_{0,\alpha}/U = \Gamma_{0}/U = 0.1$
and $\Delta \Gamma_{\alpha}=\Delta \Gamma=0.05\, U$ in our numerical
calculations.

As already stressed, the analysis is restricted to the first-order
adiabatic correction. Hence, since the current is linear in $\Omega$,
the charge pumped per cycle does not depend on the pumping rate. The
accuracy of this approximation depends on the magnitude of the
second-order corrections. Intuitively, the adiabatic approximation
becomes more accurate as the ratio $\hbar\Omega/\Gamma_0$ becomes
smaller. A closer analysis of the time derivatives of the Green's
functions induced by the adiabatic expansion reveals that the
dimensionless parameter controlling the adiabaticity is rather $\xi
={\rm max}\{\hbar\Omega/\Gamma_0,\hbar\Omega
\varepsilon_1/\Gamma_0^2\}$. Albeit the fact that the results
presented here are always valid for a sufficiently slow pumping, such
that $\xi \ll 1$, there is no simple way to estimate the accuracy of
the approximation for a given pumping rate $\Omega$. To be
quantitative, one has to evaluate the second-order correction within
the adiabatic approximation, which is a quite daunting task. Instead,
we did a rough estimate of these higher-order contributions by
studying a single representative term that appears in the second-order
Green's function. We found that it scaled with $\xi$ as predicted, up
to a numerical factor of order one.

\begin{figure}[htbp]
\begin{center}
\includegraphics[width = 8cm, angle=0]{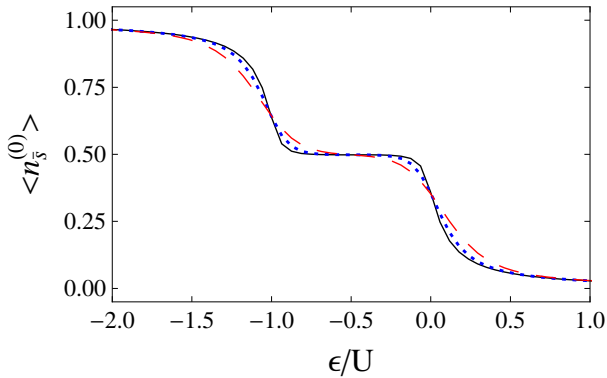}
\end{center} 
\caption{Equilibrium quantum dot occupation number $\langle
  n_{\bar{s}}^{(0)}\rangle$ as a function of the level position
  $\varepsilon$ for three values of the temperature: $k_{B}T/U = 0.01$
  (full black curve), $k_{B}T/U = 0.05$ (dotted blue curve), and
  $k_{B}T/U = 0.1$ (dashed red curve). Here $\Gamma_0=0.1\, U$ and the
  Fermi energy is set to zero, $\varepsilon_F=0$.}
\label{fig:Fig2}
\end{figure}

Figure~\ref{fig:Fig2} displays the result of the self-consistent
calculation of the zeroth order occupation $\langle n^{(0)}_{\bar{s}}
\rangle$, Eq.~(\ref{FD}), as function of the position of the resonance
$\varepsilon$ for three temperature values. Knowledge of $\langle
n^{(0)}_{\bar{s}}\rangle$ is crucial for computing the various terms
that enter in the calculation of the pumping current. As expected, the
occupation of the quantum dot increases whenever the position of any
of its two levels, $\varepsilon$ and $\varepsilon + U$, coincides with
Fermi level $\varepsilon_{F} = 0$, facilitating charge transport. For
low temperatures, this is the dominant mechanism of transport, whereas
for higher temperatures thermal fluctuations can also induce charge
transfer through the quantum dot. This explains why the features in
the curve become sharper as temperature decreases.

\begin{figure}[htbp]
\begin{center}
\includegraphics[width = 8cm,angle=0]{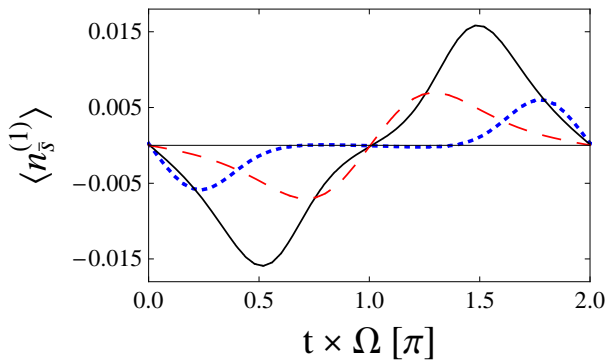}
\end{center} 
\caption{First-order correction to the quantum dot occupation number,
  $\langle n^{(1)}_{\bar{s}} \rangle$, as a function of time over a
  complete pumping cycle for three values of $\varepsilon_0$:
  $\varepsilon_0/U = -0.075$ (dotted blue curve), $\varepsilon_0/U =
  0$ (full black curve), and $\varepsilon_0/U = 0.075$ (dashed red
  curve). Temperature is $k_{B}T/U = 0.01$, $\phi_{L}= -\phi_{R} =
  \pi/2$, $\varepsilon_{1}/U = 0.05$, $\Gamma_0/U = 0.1$, and $\Delta
  \Gamma/U = 0.05$.}
  \label{fig:Fig3}
\end{figure}

The first-order correction to the quantum dot occupation number
$\langle n^{(1)}_{\bar{s}}\rangle$, also calculated self consistently
using Eq.~(\ref{FD}), is shown in Fig.~\ref{fig:Fig3} as a function of
time for several values of $\varepsilon_{0}$. It is important to
emphasize that $\langle n^{(1)}_{\bar{s}}\rangle$ is intrinsically a
time-dependent quantity and depends on the pumping parameters
dynamics, in contrast to $\langle n^{(0)}_{\bar{s}}\rangle$.  Notice
that the magnitude of $\langle n^{(1)}_{\bar{s}}\rangle$ is typically
much smaller than $\langle n^{(0)}_{\bar{s}}\rangle$.
We observe that the maximum values of $\langle n^{(1)}_{\bar{s}}
\rangle$ occur for $\varepsilon_{0} = \varepsilon_F$. When the
position of the level $\varepsilon_{0}$ deviates significantly from
$\varepsilon_{F}$, charge pumping is attenuated and the magnitude of
the current is smaller.

\begin{figure}[htbp]
\begin{center}
\includegraphics[width = 8cm,angle=0]{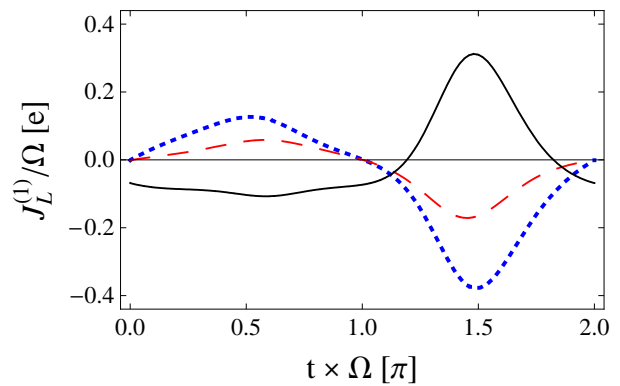}
\end{center} 
\caption{The three terms that contribute to the first-order correction
  to the pumping current as a function of time: $J^{(1){\rm
      fl}}_{\alpha} (t)$ (dotted blue curve), $J^{(1a){\rm
      dis}}_{\alpha} (t)$ (dashed red curve), and $J^{(1b){\rm
      dis}}_{\alpha} (t)$ (full black curve). Here we set
  $\varepsilon_0=0$ and take the other model parameters as in
  Fig.~\ref{fig:Fig3}.
}
\label{fig:Fig4}
\end{figure}

After computing $\langle n^{(1)}_{\bar{s}} \rangle$, the next step is
to calculate the first-order correction to the time-dependent current
$J^{(1)}_{\alpha} (t)$ given by the sum of the fluctuation term
$J^{(1){\rm fl}}_{\alpha} (t)$, Eq.~(\ref{fluctuation1}), and the
dissipation terms $J^{(1a){\rm dis}}_{\alpha} (t)$ and $J^{(1b){\rm
    dis}}_{\alpha} (t)$, Eqs.~(\ref{dissipation1a}) and
(\ref{dissipation1b}), respectively. A typical result is shown in
Fig.~\ref{fig:Fig4} where we plot the frequency independent quantity
$J^{(1)}_\alpha/\Omega$ as a function of time over a full pumping
cycle. It is important to point out that the second dissipation term,
$J^{(1b){\rm dis}}_{\alpha s} (t)$, does not contribute to the total
charge pumped per cycle since it is proportional to a total time
derivative. Consequently, its time integral over a complete pumping
cycle must vanish, a result that has been confirmed numerically. The
analysis of Fig.~\ref{fig:Fig4} reveals that these three current
terms, as $\langle n^{(1)}_{\bar{s}}\rangle$, exhibit maxima precisely
at the instants when the resonance energy level $\varepsilon(t)$
crosses the Fermi energy. In the case of Fig.~\ref{fig:Fig4}, where
$\varepsilon_0=0$, these maxima occur at $t = \pi/2\Omega$ and $t =
3\pi /2\Omega$.

\begin{figure}[htbp]
\begin{center}
\includegraphics[width = 8cm,angle=0]{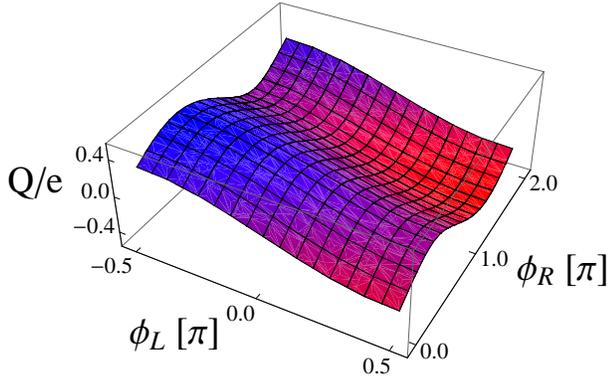}
\end{center} 
\caption{Three-dimensional graph of $Q$ as a function of $\phi_L$ and
  $\phi_R$. Temperature is $k_{B}T/U = 0.01$ while $\varepsilon_0=0$,
  $\varepsilon_{1}/U = 0.05$, $\Gamma_0/U = 0.1$, and $\Delta \Gamma/U
  = 0.05$.}
\label{fig:Fig7}
\end{figure}

There is an intuitive interpretation for the role of the pumping
parameters of our model, $\Gamma_{R,L}(t)$ and $\varepsilon(t)$, that
helps us to understand the time dependence observed above: In
Eq.~(\ref{eq:e(t)}) we fixed the phase offset of $\varepsilon(t)$ to
zero. In this situation, for $0\le t \le \tau_{\rm pump}/2$ the
resonance energy $\varepsilon$ decreases with time. As a consequence,
during this half pumping period $\langle n_s \rangle$ increases with
time, which corresponds to loading negative charge into the quantum
dot. In this time interval, the sign of the pumping current depends on
the phase difference between $\phi_R$ and $\phi_L$. The situation is
reversed for $\tau_{\rm pump}/2 \le t \le \tau_{\rm pump}$. Figure
\ref{fig:Fig7} shows the three-dimensional plot of the charge pumped
per cycle $Q$ as a function of both $\phi_R$ and $\phi_L$. Consistent
with the reasoning presented above, having $\phi_L$ and $\phi_R$ in
anti-phase favors larger values of $|Q|$. In particular, we find two
maximum values of $|Q|$, one at $\phi_L=\pi/2$ and $\phi_R=3\pi/2$,
and the other at $\phi_L=-\pi/2$ and $\phi_R=\pi/2$. The location of
these maxima shows no dependence on any of the model parameters,
provided $\varepsilon_1\neq 0$. In this limit case, there are only two
active pumping parameters, $\Gamma_R$ and $\Gamma_L$, and the
dependence of $Q$ on the $\phi_R$ and $\phi_L$ is the same as in the
non-interacting case.\cite{brouwer98} Since we are interested in
maximizing $|Q|$, in the remaining of this paper we take
$\varepsilon_1\neq 0$ and $\phi_L=-\phi_R=\pi/2$.

We are now ready to study the dependence of $Q$ on $V_g(t)$, related
to $\varepsilon_0$ and $\varepsilon_1$, as well as on the dot-lead
couplings, represented in our model by $\Gamma_0$ and $\Delta \Gamma$.

\begin{figure}[htbp]
\begin{center}
\includegraphics[width = 8cm,angle=0]{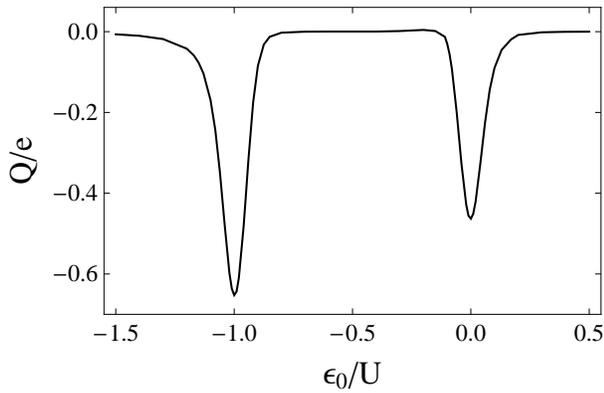}
\end{center} 
\caption{Charge pumped per cycle as a function of the level position
  $\varepsilon_{0}$ for $\varepsilon_{1}/U = 0.05$, $k_BT/U=0.01$, 
  $\phi_{L}= -\phi_{R} = \pi/2$, $\Gamma_0/U = 0.1$, and $\Delta \Gamma/U
  = 0.05$. The charge is measured in units of the
  electron charge $e$.}
\label{fig:Fig5}
\end{figure}

In Fig.~\ref{fig:Fig5} we show the charge pumped per cycle $Q$
calculated as a function of $\varepsilon_{0}$. Charge pumping is
enhanced whenever a quantum dot resonance, $\varepsilon_{0}$ or
$\varepsilon_{0} + U$, crosses the Fermi level, resulting in the two
peaks of Fig.~\ref{fig:Fig5}.

\begin{figure}[htbp]
\begin{center}
\includegraphics[width = 8cm,angle=0]{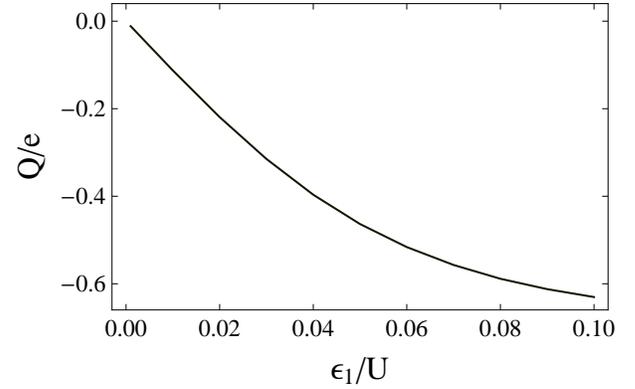}
\end{center} 
\caption{Charge pumped per cycle as a function of the resonance
  oscillation amplitude $\varepsilon_{1}$ for $\varepsilon_{0} = 0$,
  $\Gamma_0/U=0.1$, $\Delta \Gamma/U=0.05$, $k_BT/U=0.01$, and
  $\phi_{L}= -\phi_{R} = \pi/2$. The charge is measured in units of
  the electron charge $e$.}
\label{fig:Fig6}
\end{figure}

Figure \ref{fig:Fig6} shows the dependence of $Q$ on $\varepsilon_1$.
We consider one of the situations of maximum pumping, namely,
$\phi_L=-\phi_R=\pi/2$ and $\varepsilon_0=\varepsilon_F=0$.  In this
case, $|Q|$ increases monotonically with $\varepsilon_1$. We caution
that once $\varepsilon_1$ exceeds $\Gamma_0$, it is necessary to check
whether $\xi \ll 1$, so that the adiabatic approximation still holds.
Hence, increasing $\varepsilon_1$ might not be advantageous whenever
it is necessary to reduce $\Omega$. Figure \ref{fig:Fig6} also shows
that $Q$ vanishes when $\varepsilon_{1}$= 0, as expected for a
two-parameter adiabatic pump that occur for
$\phi_L=-\phi_R=\pi/2$.\cite{thouless,brouwer98}

\begin{figure}[htbp]
\begin{center}
\includegraphics[width = 8cm,angle=0]{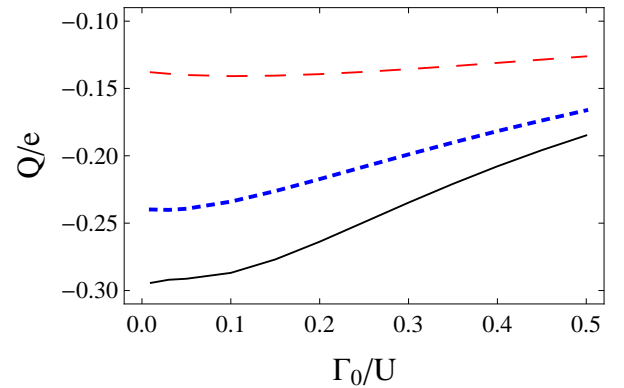}
\end{center} 
\caption{Charge pumped per cycle as a function of $\Gamma_0$ for
  different values of temperature: $k_BT/U=0.05$ (full black curve),
  $k_BT/U=0.1$ (dotted blue curve), and $k_BT/U=0.2$ (dashed red
  curve). Here $\varepsilon_0=\Gamma$, $\varepsilon_1/U=0.05$, $\Delta
  \Gamma/\Gamma_0=1$, and $\phi_{L}= -\phi_{R} = \pi /2$.}
\label{fig:QvsGammaall}
\end{figure}

We now address the dependence of $Q$ on $\Delta \Gamma$ and
$\Gamma_0$. To be quantitative, we now also keep $T \gg T_K$ for the
sake of the validity of our approximation. To maximize pumping, we
find that it is advantageous to decrease $T_K$ by taking
$\varepsilon_0 \neq 0$ rather than increasing $T$. As before, we
consider $\phi_L=-\phi_R=\pi/2$.  Due to the time derivatives
appearing in the Green's function expressions, several terms in
Eqs.~(\ref{fluctuation1}) and (\ref{dissipation1a}) are proportional
to $\Delta \Gamma$. Indeed, we find that $Q$ is roughly linear in
$\Delta \Gamma$ for several values of $\varepsilon_1 \le \Gamma_0$.
Figure \ref{fig:QvsGammaall} shows $Q$ versus $\Gamma_0$ for three
temperature values. Due to the fact that $k_BT_K \le \sqrt{\Gamma U
  /2}e^{-\pi/2}$ for $\varepsilon \approx \Gamma_0$, our approximation
scheme breaks down as $\Gamma_0$ is increased and $T_K$ reaches $T$.

\begin{figure}[htbp]
\begin{center}
\includegraphics[width = 8cm,angle=0]{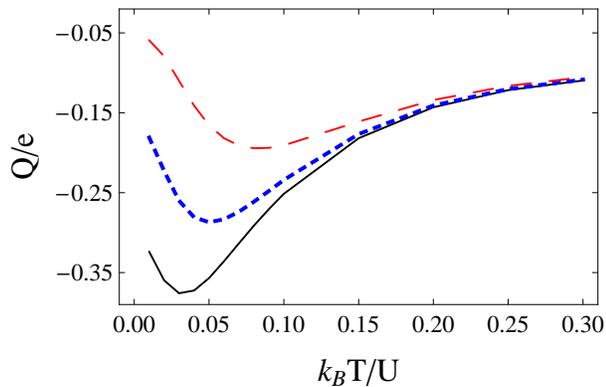}
\end{center} 
\caption{Charge pumped per cycle as a function of temperature for
  $\varepsilon_0/U=0.075$ (full black curve), $\varepsilon_0/U=0.1$
  (dotted blue curve), and $\varepsilon_0/U=0.15$ (dashed red
  curve). For all curves $\varepsilon_1/U=0.05$, $\Gamma_0/U=0.1$,
  $\Delta \Gamma/U=0.1$, and $\phi_{L}= -\phi_{R} = \pi /2$.}
  \label{fig:QvsT}
\end{figure}

Figure \ref{fig:QvsT} shows $Q$ as a function of temperature for three
values of the resonance energy $\varepsilon_0$ with $\varepsilon_1$
kept fixed. The temperatures for which we observe the largest values 
of $|Q|$ scale with $\varepsilon_0$. We also find
that by decreasing $|\varepsilon_0|$ the maximum of $|Q|$
increases. Unfortunately, since our results are only valid for $T\gg
T_K$, we cannot freely vary $\varepsilon_0$.

\begin{figure}[htbp]
\begin{center}
\includegraphics[ width = 8cm,angle=0]{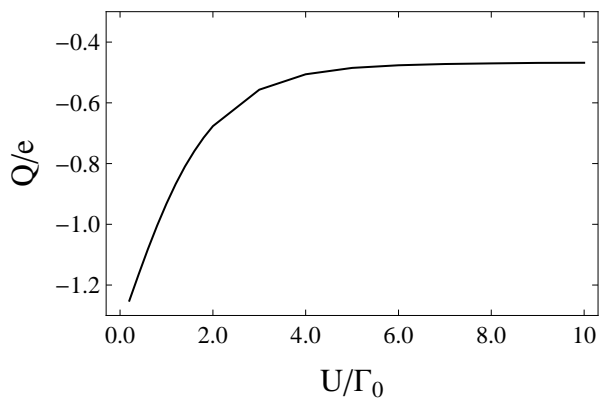}
\end{center} 
\caption{Charge pumped per cycle as a function of $U/\Gamma_0$ for
  $\varepsilon_0=\Gamma_0$, $\varepsilon_1=\Gamma_0/2$, and $k_B
  T=\Delta \Gamma =\Gamma_0$.}
\label{fig:QvsU}
\end{figure}

Finally, let us address the dependence of $Q$ on the charging energy
$U$. Our results are summarized in Fig.~\ref{fig:QvsU}. A large
interval range for $U$ is displayed to best illustrate the pumped
charge dependence on this parameter. We observe that pumping is
largely enhanced for small values of $U$. When $U$ becomes comparable
to $\Gamma$ the system departs from the Coulomb blockade regime.

\section{Conclusions}
\label{sec:conclusions}

In conclusion, we have investigated adiabatic charge pumping through
quantum dots in the Coulomb blockade regime. We specifically studied
the impact of Coulomb interaction on the pumping current amplitude for
the \emph{finite-U} Anderson model, in contrast to previous works that
treated the infinite-$U$ case.\cite{Splettstoesser05}

We have derived a general expression for the adiabatic pumping current
that is proportional to the instantaneous Green's function of the
dot. This formula was then applied to compute the time dependence of
the total charge pumped per cycle through the dot. This allowed us to
analyze several aspects of experimental relevance, such as the
dependence of the pumped charge on temperature and on the phase
difference between time-dependent perturbations.

We find that, within the adiabatic regime, there is a large range of
parameters that can be used to maximize the charge pumped per
cycle. For this purpose, we find that it is advantageous to: (i) Tune
the back gate voltage to pump with the QD in resonance with the Fermi
energy in the leads; (ii) maximize the pumping amplitude $\Delta
\Gamma$ and, possibly, $\varepsilon_1$ as well; (iii) minimize
temperature.

We were not able to find a set of parameter values that gives one unit
of charge $e$ per pumping cycle within the parameter ranges allowed by
our approximations. We do not discard such interesting possibility,
but our investigations hint that it may only be possible for very
particular pulse formats, not necessarily sinusoidal, and within a
narrow parameter interval. The possibility of spin pumping and the
consideration of the double-dot case are under investigation and will
be reported soon.

\section*{Acknowledgments}

We acknowledge partial financial support from the Brazilian funding
agencies CNPq and FAPERJ. This work was also made possible by the
American Physical Society International Travel Grant Award.

Note: On the process of completing this study, we became aware of
Ref.~\onlinecite{Winkler09} that deals with a similar problem using
the diagrammatic real-time approach.




\begin{thebibliography}{99}
\bibitem{footnote1} Although this demonstration has been made for the
  retarded component, it also applies for the lesser and greater ones.
\end{thebibliography}
\end{document}